\documentclass[a4paper,11pt]{article}
\usepackage{amsmath}
\usepackage{amssymb}
\usepackage{graphicx}
\usepackage[sort]{cite}
\usepackage[T1]{fontenc}
\usepackage[a4paper,left=1.0in, right=1.0in,top=1.1in, bottom=1.2in]{geometry}
\usepackage[affil-it]{authblk}
\usepackage{color}
\usepackage{subcaption}
\usepackage[colorlinks=true,linktocpage=true,linkcolor=blue,citecolor=blue]{hyperref}

\newcommand{\be}{\begin{equation}}
\newcommand{\ee}{\end{equation}}
\def\dd{\text{d}}

\def\q{{\boldsymbol q}}
\def\r{{\boldsymbol r}}
\def\x{{\bf x}}
\def\y{{\bf y}}
\def\b{{\bf b}}
\def\z{{\boldsymbol z}}
\def\pT{p_t}

\begin{document}

\title{\Large \bf Towards tomography of quark-gluon plasma using double inclusive forward-central jets in Pb-Pb collision}

\author{Michal Deak and Krzysztof Kutak}
\affil{Instytut Fizyki Jadrowej, Radzikowskiego 152, 31-342 Krak\'ow, Poland}
\author{Konrad Tywoniuk}
\affil{Theoretical Physics Department, CERN, 1211 Geneva 23, Switzerland}

\maketitle

\begin{abstract}
We propose a new framework, merging High Energy Factorization with final-state jet quenching effects due to interactions in a quark-gluon plasma, to compute di-jet rates at mid- and forward-rapidity. It allows to consistently study the interplay of initial-state effects with medium interactions, opening the possibility for understanding the dynamics of hard probes in heavy-ion collisions  and the QGP evolution in rapidity.

\end{abstract}

\section{Introduction}
The Large Hadron Collider (LHC) offers unprecedented possibilities to study properties of nuclear matter in extreme conditions.
One of the paramount results obtained at LHC, and earlier at the Relativistic Heavy Ion Collider (RHIC), is the strong evidence of a state of matter called quark-gluon plasma (QGP) in ultrarelativistic heavy-ion collisions 
through its strong quenching of perturbative probes, such as heavy-quark and QCD jet production.
In particular, jets are excellent tools in the study of the properties of the QGP due to their coupling to the deconfined plasma degrees of freedom, and they allow one to study new aspects of QCD dynamics, for reviews see \cite{Mehtar-Tani:2013pia,Blaizot:2015lma}.
In particular, recent measurements of single-inclusive jet modifications \cite{Abelev:2013kqa,Aad:2014bxa,Khachatryan:2016jfl} and di-jet asymmetry \cite{Aad:2010bu,Chatrchyan:2011sx,Chatrchyan:2012nia} have also attracted interest from theory \cite{CasalderreySolana:2012ef,Blaizot:2013hx}; see also \cite{Qin:2015srf} and the references therein. On theoretical grounds, the latter phenomenon was understood to be a consequence of fluctuations of final-state jet energy loss, leading to an on average dispersion of the energy difference. Because the typical medium scales, related through multiple scattering to the jet quenching coefficient $\hat q$, are modest compared to the jet energies, the di-jet angular correlation remains practically unmodified compared to its vacuum baseline. The studies performed so far were limited to a strictly back-to back configuration of the di-jets at mid-rapidity well suited for calculations within collinear factorization.

It is worth pointing out that the di-jet azimuthal distribution away from the back-to-back configuration is dominated by vacuum effects related to initial-state (space-like) emissions. It would therefore be interesting to study in greater detail the interplay of these contributions and the novel, final-state modifications arising in heavy-ion collisions. This was first addressed in the context of di-jet azimuthal decorrelation due to in-medium transverse momentum broadening \cite{Mueller:2016gko,Mueller:2016xoc}. In the current study, we focus instead on effects related to medium-induced radiative energy loss. Additionally, we allow for jet production at more forward rapidities than considered in heavy ion collisions so far. This opens for an interesting exploration of the nuclear wave-function in tandem with medium effects.
Such a framework would be of general interest for providing a consistent cross-referencing of observables calculated across various hadronic colliding systems, especially proton-nucleus and nucleus-nucleus collisions. Finally, it makes new use of jets as tomographic probes of the rapidity profile of the QGP.

The framework which allows us to study from the first principles the full angular dependence of decorrelations of forward-central jet configuration in vacuum is the hybrid High Energy Factorization \cite{Catani:1990eg,Deak:2009xt,Kutak:2012rf}. 
In this approach, the kinematics is treated exactly from the outset, the matrix elements are calculated with one  the incoming parton's momenta (carrying low longitudinal momentum fraction of parent hadron) off-shell and one on-shell (carrying large fraction of parent hadron). The incoming off-shell parton carries transversal momenta which allow for the decorrelation of final-state jets. 
In this approach, the kinematics is treated exactly from the outset, the matrix elements are calculated with one of the incoming parton's momenta (carrying small longitudinal momentum fraction of the parent hadron momentum) off-shell and one on-shell (carrying large fraction of the parent hadron momentum). The incoming off-shell parton carries transversal momenta which allow for decorrelation of final state jets. 
In this approach, in order to calculate cross sections, the matrix element needs to be convoluted with the transverse momentum dependent (TMD) parton density function as well as with standard PDF parametrizing partons carrying large longitudinal momentum fraction of parent hadron (in our case a typical $x$ on the 'projectile' side is $10^{-1}$ ). In particular the transversal momentum dependent PDF could be provided by the BFKL \cite{Kuraev:1977fs,Balitsky:1978ic,Kuraev:1976ge} equation, when the longitudinal momenta are small but the system is sufficiently dilute to obey linear dynamics, or by the KMRW framework, when the longitudinal momenta are moderate as considered here sue up (in our case a typical on the 'target' side is $x\simeq 10^{-3}$). The latter framework allows for a transformation of the collinear gluon density to the TMD PDF by the so called Sudakov resummation.

In the more extreme situations, i.e. when the parton densities are probed at low-$x$, one needs to account for eventual saturation effects \cite{Gribov:1984tu}. This complicates the factorization formula since, besides taking into account the dipole gluon density which is a solution of the Balitsky-Kovchegov equation \cite{Balitsky:1995ub,Kovchegov:1999yj,Kovchegov:1999ua}, one also needs to take into account the Weizs\"acker-Williams gluon density \cite{Dominguez:2011wm}. 

By combining the hybrid HEF with final-state rescatterings in a hot and dense medium created during nucleus-nucleus collisions,  we propose a framework which allows for well controlled study of the full azimuthal dependence of the cross section and for investigations of the longitudinal structure of QGP at the same time.
Encouraged by the success of HEF in describing various data \cite{Bury:2016cue,Kutak:2016mik} we shall apply it to central-forward di-jet production in heavy-ion collisions by including effects relevant for jets passing through a hot and dense QCD medium into the HEF Monte Carlo generator KaTie~\cite{vanHameren:2016kkz}. We argue that di-jet observables in HEF are more suitable to study rapidity/rapidity-azimuthal structure of the quark-gluon plasma formed in a heavy-ion collisions. Owing to the factorization of soft, medium-induced radiation from the hard vertex, the final-state modifications are then implemented as energy-loss probabilities affecting final-state particles \cite{Baier:2001yt,Salgado:2003gb}.

Recently, the importance of jet substructure fluctuations on the di-jet asymmetry and the generic energy-loss mechanism was pointed out \cite{Milhano:2015mng,Mehtar-Tani:2017ypq}. 
In this exploratory study, we will, however, not consider further details of jet fragmentation. We structure the paper in the following way. We present the details of the framework and implementation of medium effects in Sect.~\ref{sec:framework}. Numerical results for the production  of central-forward di-jets in heavy-ion collisions at the LHC are presented in Sect.~\ref{sec:numerics}, and finally we discuss our results and provide a brief outlook in Sect.~\ref{sec:conclusions}.

\section{General framework and implementation of medium effects}
\label{sec:framework}
In order to calculate the cross section for the double inclusive jet production with medium effects included one needs to generalize the vacuum framework, which in our case refers to proton-proton collisions. The generalization is two-fold: 
\begin{itemize}
\item replacement of collinear PDF by nPDF and by replacement of TMD by nTMD;
\item acconting for energy loss.
\end{itemize}
The formula for hybrid HEF in dilute-dilute scattering reads \cite{Deak:2009xt,Iancu:2013dta}:
\be
  \frac{d\sigma_{acd}}{dy_1dy_2dp_{t1}dp_{t2}d\Delta\phi} 
  =
  \frac{p_{t1}p_{t2}}{8\pi^2 (x_1x_2 S)^2}
  |\overline{{\cal M}_{ag^*\to cd}|}^2
  x_1 f_{a/A}(x_1,\mu^2)\,
  {\cal F}_{g/B}(x_2,k_t^2,\mu^2)\frac{1}{1+\delta_{cd}}\,,
  \label{eq:cs-main}
\ee
with  $k_t^2 = p_{t1}^2 + p_{t2}^2 + 2p_{t1}p_{t2} \cos\Delta\phi$, and
$$
  x_1 = \frac{1}{\sqrt{S}} \left(p_{t1} e^{y_1} + p_{t2} e^{y_2}\right)\,,
  \qquad
  x_2 = \frac{1}{\sqrt{S}} \left(p_{t1} e^{-y_1} + p_{t2} e^{-y_2}\right)\,,
  \label{eq:x1x2}
$$
where $|\overline{{\cal M}_{ag^*\to cd}|}$ is the hard matrix element for scattering of on-shell parton $a$ off a space-like gluon\footnote{The contribution from off-shell quarks for the studied jet configuration is negligible.} to partons $c$ and $d$. The matrix elements can be found in \cite{Kutak:2012rf} or evaluated using helicity methods \cite{vanHameren:2012if}. The distribution ${\cal F}(x,k_t^2,\mu^2)$ is an unintegrated gluon density parametrizing the partonic content of a hadron carrying a small longitudinal momentum fraction $x$ of the parent hadron and some transverse momentum $k_t$. This PDF depends, in general, on some factorization scale $\mu$. It is obtained via the application of the KMRW framework, i.e., by performing a resummation of soft gluons using the Sudakov form factor \cite{Kimber:1999xc,Watt:2003vf}. The formulation is such that, upon integration over the transversal momentum up to hard scale $\mu$, one recovers the collinear gluon density. The function $  x_1 f_{a/A}(x_1,\mu^2)$ is a collinear PDF characterizing partons carrying large longitudinal momentum fractions and probed at the hard scale $\mu$.

In order to calculate the cross section for propagation of di-jets through medium produced in heavy-ion collision we need to extend the HEF framework to account for the energy loss of jets traversing the medium. For high-$\pT$ jets, one can safely assume the dominance of radiative processes from medium-induced bremsstrahlung. The emission spectrum of medium-induced gluons can be factorised from the hard process and is given by
\begin{align}
\label{eq:bdmps-spectrum}
\omega \frac{d I_{\scriptscriptstyle R}(\chi)}{ d \omega} &= \frac{\alpha_s\,C_R}{\omega^2}  2\text{Re} \int^{\chi \omega}\frac{d^2 \q}{(2\pi)^2 }\int_0^\infty d t' \int_0^{t'} d t \int d^2 \z \exp\left[ - i \q \cdot \z - \frac{1}{2} \int_{t'}^\infty\dd s \, n(s) \sigma(\z) \right]\nonumber\\
&\times  \partial_{\z}\cdot \partial_{\y} \left[ \mathcal{K}(\z,t'; \y ,t |\omega) -  \mathcal{K}_0(\z,t'; \y ,t |\omega)\right]_{\y = 0} \,,
\end{align}
in terms of the gluon energy $\omega$ and transverse momentum $\q$ with respect to the jet axis \cite{Baier:1996sk,Baier:1996kr,Zakharov:1996fv,Zakharov:1997uu,Wiedemann:2000za}. The spectrum Eq.~(\ref{eq:bdmps-spectrum}) is, in fact, independent of the jet direction. The function
\be
\mathcal{K}(\z,t'; \y ,t |\omega) = \int_{\r(t) = \y}^{\r(t') = \z} \mathcal{D}\r\, \exp \left\{\int_t^{t'} \dd s\left[ i\frac{\omega}{2}\dot \r^2 - \frac{1}{2} n(s) \sigma(\r) \right] \right\} \,, 
\ee
is the solution to a 2D Schr\"odinger equation describing rescattering in the medium governed by a medium gluon density $n(s)$ along the path of propagation. Finally, $\sigma(\r)$ is related to the medium interaction potential. In Eq.~(\ref{eq:bdmps-spectrum}), we have explicitly subtracted the vacuum contribution $\mathcal{K}_0 \equiv \lim_{n(s) \to 0} \mathcal{K} $, which corresponds to the free gluon Green function. Further vacuum showering is not considered, in accordance with Eq.~(\ref{eq:cs-main}). The spectrum is proportional to the colour factor of the projectile, for a fast quark (gluon) $C_{\scriptscriptstyle R} = C_F$ ($C_{\scriptscriptstyle R} = N_c$) and is a function of the factor $\chi$, which parameterizes the angular range of the emitted gluons. In Eq.~\ref{eq:bdmps-spectrum}, $\chi=\sin\Theta$ where $\Theta$ is the angle between jet axis and radiated emission. We will assume $\chi =1$, corresponding to gluons' emitted angles $\leq \pi/2$.

Due to the steeply falling spectrum of hard particles, energy loss will be dominated by multiple emissions of soft gluons \cite{Baier:2001yt}. Due to the typical short formation time, this warrants a description in terms of multiple independent emissions; for recent improvements see \cite{Arnold:2015qya}. Since we are interested in computing the energy emitted off a high-energy projectile, we will only resum primary emissions and neglect, for the moment, further cascading. This can be further justified by the lack of cone definition in our setup. Hence, the probability of emitting a total energy $\epsilon$ can be written as 
\be
P_{\scriptscriptstyle R}(\epsilon) =\Delta(L) \, \sum_{n=0}^\infty \,\frac{1}{n!} \, \prod_{i=1}^n \, \int_0^L\dd t   \int \dd \omega_i   \frac{\dd I_{\scriptscriptstyle R}(\chi)}{\dd \omega_i \dd t  }  \, \delta\left(\epsilon-\sum_{i=1}^n\omega_i\right) \,,
\ee
with
\be
\Delta(L)\equiv \exp\left( -\int_0^L\dd t \int_0^\infty \dd \omega\, \frac{\dd I_{\scriptscriptstyle R}(\chi)}{\dd \omega \dd t} \right)
\ee
being the Sudakov form factor that represents the probability of not radiating between 0 and L. Concretely, we will use the numerical implementation utilized in \cite{Salgado:2003gb}.

\begin{figure}[t!]
\centering
\begin{subfigure}[t]{.23\textwidth}
\includegraphics[scale=.52]{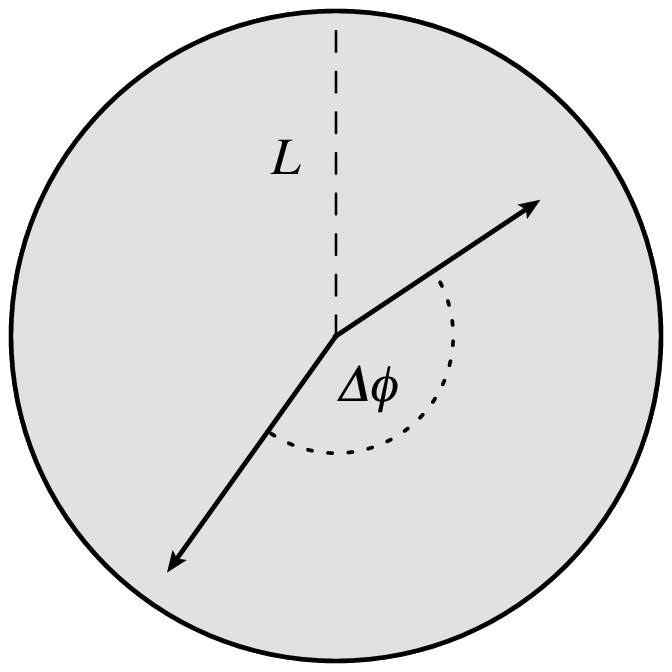}
\caption{}
\label{fig:illustrationsa}
\label{fig:simplMptc}
\end{subfigure}
\begin{subfigure}[t]{.6\textwidth}
\centering
\includegraphics[scale=.52]{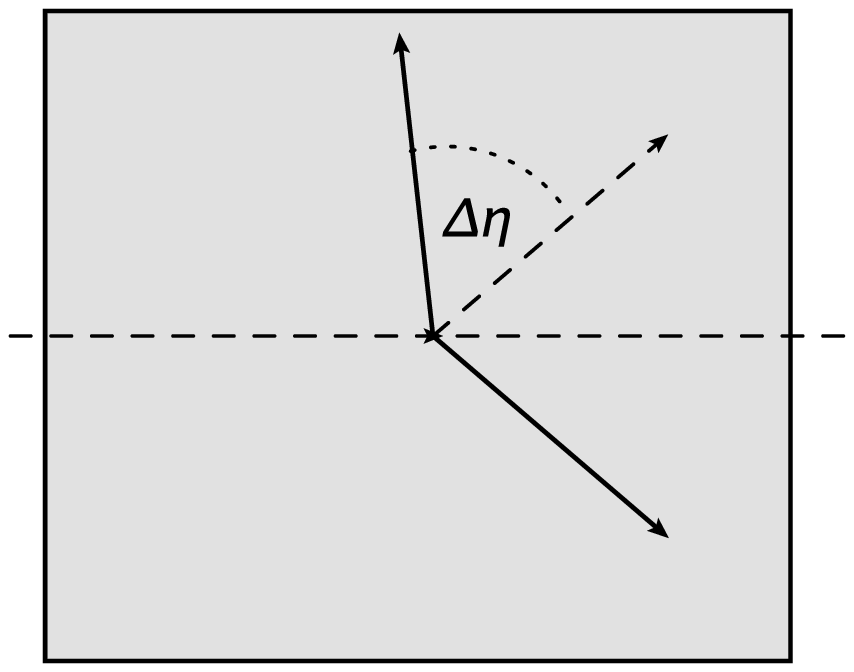}
\caption{}
\label{fig:illustrationsb}
\end{subfigure}
\caption{\small{Illustration of jets passing through the nuclear medium. Length the jets pass through the medium $L$. (a) The azimuthal cross section of the nuclear medium. (b) Longitudinal cross section of the nuclear medium.}}
\label{fig:illustrations}
\end{figure}

We employ  standard parametric estimates to argue that the timescale for the hard process is much smaller than the timescales related to soft, medium-induced radiation in the final state. Furthermore, momentum broadening effects are neglected due to the smallness of the medium parameters compared to the typical jet energies which result only in very small deflection angles.
This allows us to generalize the HEF formula as 
\be
\frac{\dd \sigma}{d y_1 d y_2 d p_{t1} d p_{t2} d\Delta\phi} = \sum_{a,c,d} \int_0^\infty d \epsilon_1 \int_0^\infty d \epsilon_2 \, P_a(\epsilon_1) P_g(\epsilon_2)\, \left. \frac{d \sigma_{acd}}{d y_1 \dd y_2 d p'_{t1} d p'_{t2} d\Delta\phi} \right|_{\substack{ p'_{1t} = p_{1t} + \epsilon_1\\p'_{2t} = p_{2t} + \epsilon_2}} \,,
\label{eq:factvm}
\ee
where the Pb-Pb vacuum cross section is given by
\be
  \frac{d\sigma_{acd}}{dy_1dy_2dp_{t1}dp_{t2}d\Delta\phi} 
  =
  \frac{p_{t1}p_{t2}}{8\pi^2 (x_1x_2 S)^2}
  |\overline{{\cal M}_{ag^*\to cd}|}^2
  x_1 f^{Pb}_{a/A}(x_1,\mu^2)\,
  {\cal F}^{Pb}_{g/B}(x_2,k_t^2,\mu^2)\frac{1}{1+\delta_{cd}}\,,
  \label{eq:cs-main}
\ee
Equation (\ref{eq:factvm}) accounts for nuclear effects in the partonic content of nuclei as well as energy loss of the final-state jet particles (Fig.~\ref{fig:illustrations}). We stress that the formula above is a conjecture and assumes factorization of vacuum emissions and medium rescatterings. As explained there are indications that the formula can be justified when the jets are hard, i.e. the medium modifies their properties slightly. Furthermore the particular choice of thge factorization scale allows one to separate initial-state emissions from final-state ones. 

We will apply the harmonic approximation, consisting in writing $n(s) \sigma(\r) \approx \hat q(s) \r^2/2$, as a simple model for interactions in the QGP. 
One of the crucial elements of the formula for incorporating the medium effects comes from the transport coefficient $\hat q$. 
Assuming a thermalized QGP, it is associated to the local temperature and parametrically $\hat q \sim g^4 T^3$, where $g$ is the in-medium coupling. 
In our studies we use a model linking it with the energy  density described in \cite{Baier:2002tc}; see also \cite{Renk:2006sx}. 
It reads
\begin{equation}\label{eq:qhat}
{\hat q}=2\,K\,\varepsilon^{3/4} \,,
\end{equation}
where $K$ is a constant quantifying the deviation from expectations in a weakly coupled QGP. The energy density $\varepsilon$ is parameterized according to the data of bulk particle production, and it reads
\begin{equation}
\varepsilon=\varepsilon_\text{tot}W\left({\bf x},{\bf y};{\bf b}\right)H(\eta)
\end{equation}
where $\epsilon_\text{tot}$ is a free parameter \cite{Renk:2006sx,Nonaka:2006yn}. 
We have updated the model in order to describe the rapidity distribution of particle production at LHC, as
\begin{equation}
\label{eq:modelrap}
H(\eta)=\frac{1}{\sqrt{2\pi}\left(a_1\,b_1-a_2\,b_2\right)}\left[ a_1 \,e^{-|\eta|^2/\left(2\,b^2_{1}\right)} - a_2\, e^{ - |\eta|^2 / \left( 2\,b^2_{2}\right)} \right] ,
\end{equation}
with fitting parameters $a_1$, $b_1$, $a_2$ and $b_2$ \cite{Abbas:2013bpa}. Since the distribution (\ref{eq:modelrap}) is normalized, $\epsilon_\text{tot}$ corresponds to the total energy density distributed in the whole rapidity range. Finally, we assume a simplified geometry of the QGP, having all particles traversing the same length $L$ in the medium so that $\hat q$ is only a function of the rapidity; see Fig.~\ref{fig:illustrationsa}. This amounts to putting $W(\x,\y;\b)\to 1$ and neglect the sampling over production points and impact parameters. We choose a realistic value of $L = 5$ fm.

\begin{figure}[t!]
\centering
\includegraphics[scale=.4]{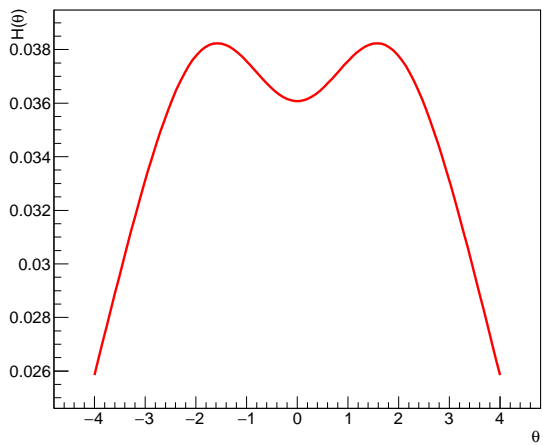}
\caption{Profile of the function $H(\eta)$.} 
\label{fig:Hmodel}
\end{figure}

In the numerical calculations we have used the following values of the parameters. First, we fix $K=1$, demanding that the value of $\hat q$ at mid-rapidity corresponds to $1$ GeV$^2$/fm. This means that $\epsilon_\text{tot}\approx 143$ GeV/fm$^3$. Varying the parameter $K$ allows us to scan a range of realistic values for $\hat q$. The remaining parameters we fit to the data on charged particles in $0-5 \%$ central collisions \cite{Abbas:2013bpa}, giving  $a_1=2108.05$, $b_1=3.66935$, $a_2=486.368$ and $b_2=1.19377$. The resulting shape of the $H(\eta )$ function is plotted in Fig.~\ref{fig:Hmodel}.

A realistic energy-loss probability distribution $P\left(\xi,r\right)$ \cite{Wiedemann:2000za}, where $\xi=\epsilon/\omega_c$ with $\omega_c={\hat q}L^2/2$ and $r={\hat q}L^3/2$, contains two components: a discrete and a continuous component,
\begin{equation}\label{eq:probC1C2}
P\left(\xi,r\right)=C_1\,\delta\left(\xi\right)+C_2\,D\left(\xi,r\right)\,.
\end{equation}
The coefficient $C_1$ gives the probability that no suppression occurs. The function $D\left(\xi,r\right)$ describes the continuous component of the probability distribution. 
With the parameters chosen above we find that $\omega_c \approx 62.5 $ GeV and $r \approx 1560$.
In the implementation of the probability distribution, with $\alpha_s =1/3$ as default, for a given event, we first generate a random number $C_r$ from $0$ to $1$. If $C_r<C_1$, then no suppression in the medium occurred, $\xi=0$ and the weight coming from the medium correction is $w_M=1$; therefore, the total weight is $w_T=w_M\,w=w$, where $w$ is the original weight for the event in the vacuum. If $C_r>C_1$, one can employ the Metropolis algorithm to generate $\xi$ according to the distribution $D\left(\xi,r\right)$, then $w_M=1$, or generate $\xi$ according to a simple distribution with a corresponding weight $w_S$, then $w_M=D\left(\xi,r\right)$ and $w_T=w_S\,w_M\,w$.
\begin{figure}[t!]
\centering
\begin{subfigure}[t]{.48\textwidth}
\includegraphics[scale=.42]{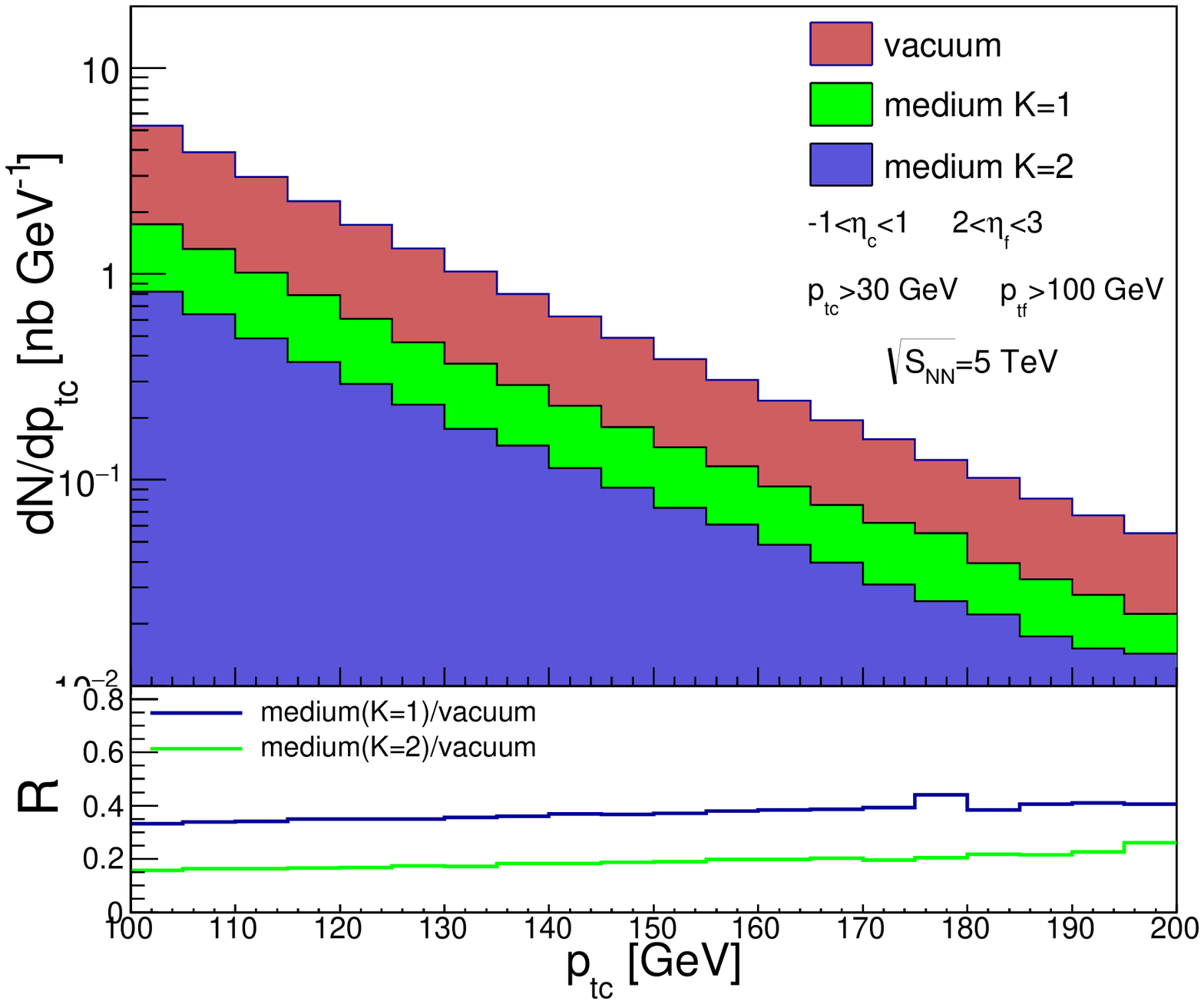}
\caption{}
\label{fig:simplMptc}
\end{subfigure}
\begin{subfigure}[t]{.48\textwidth}
\centering
\includegraphics[scale=.42]{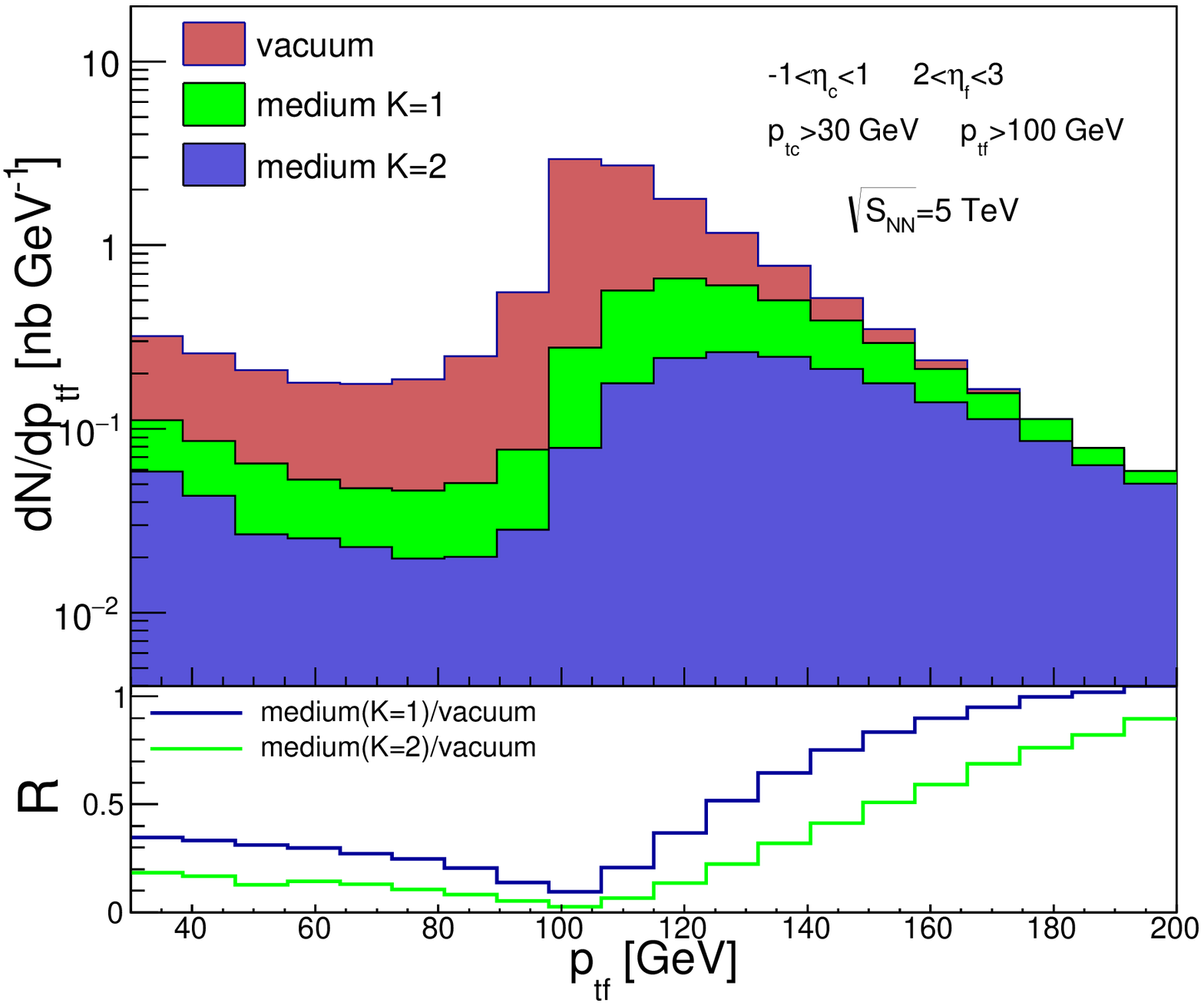}
\caption{}
\label{fig:simplMptf}
\end{subfigure}
\caption{\small{(a) On top: Central jet ${p_{t}}_c$ before and after jets pass through the medium. (b) On top:  Forward jet ${p_{t}}_f$ before and after the jets pass through the medium. The red histogram represents the ${p_{t}}_{c/f}$ of the jet without passing the medium. The light green histogram represents the ${p_{t}}_{c/f}$ spectrum of a jet quenched by the medium with constant $K=1$~\eqref{eq:qhat}. The blue histogram represents the ${p_{t}}_{c/f}$ spectrum of a jet quenched by the medium with $K=2$. On bottom: ratios of the histograms.}}
\end{figure}

\section{Numerical results}
\label{sec:numerics}
\begin{figure}[t!]
\centering
\begin{subfigure}[t]{.48\textwidth}
\includegraphics[scale=.42]{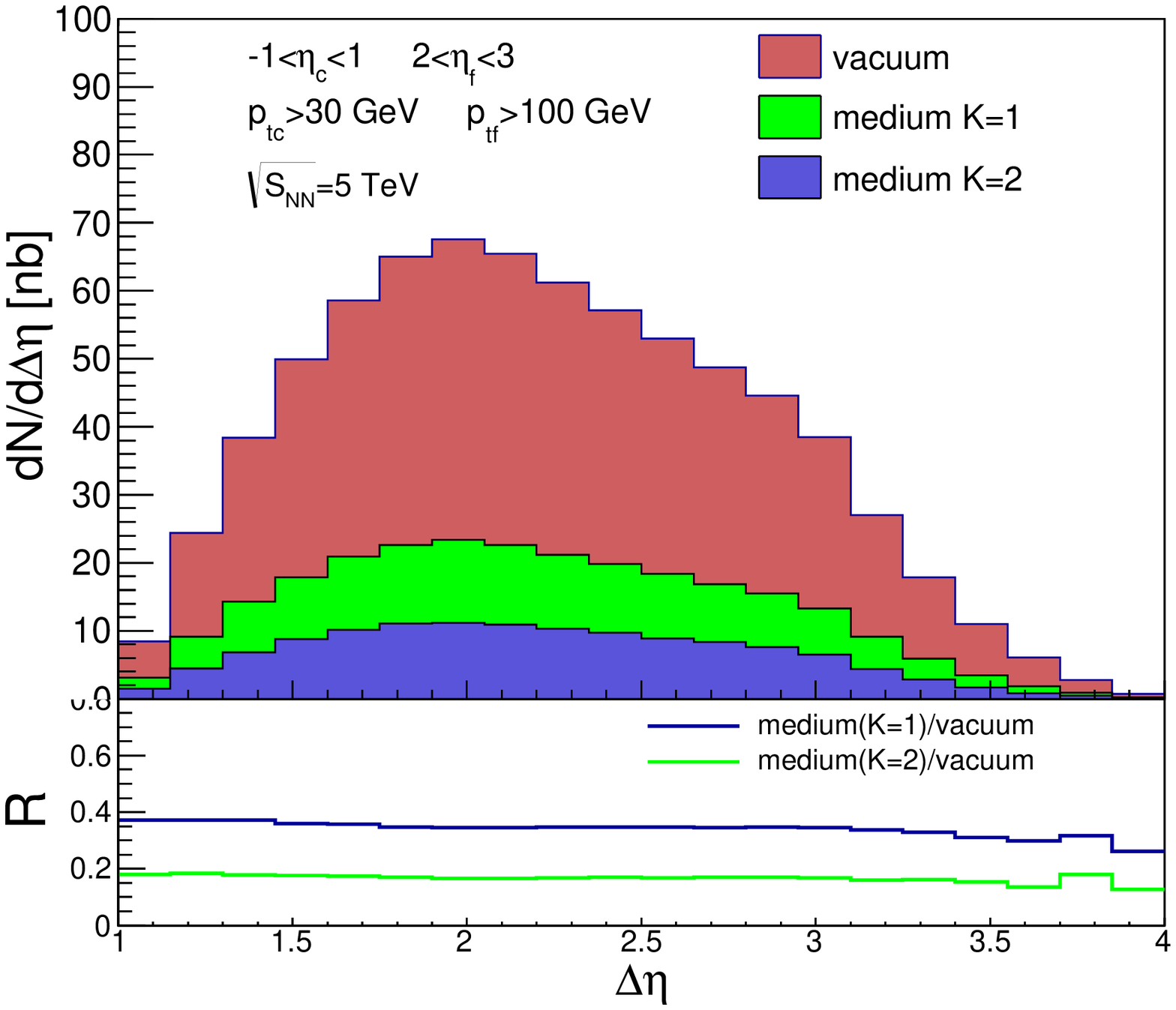}
\caption{}
\label{fig:simplMdeta}
\end{subfigure}
\centering
\begin{subfigure}[t]{.48\textwidth}
\includegraphics[scale=.42]{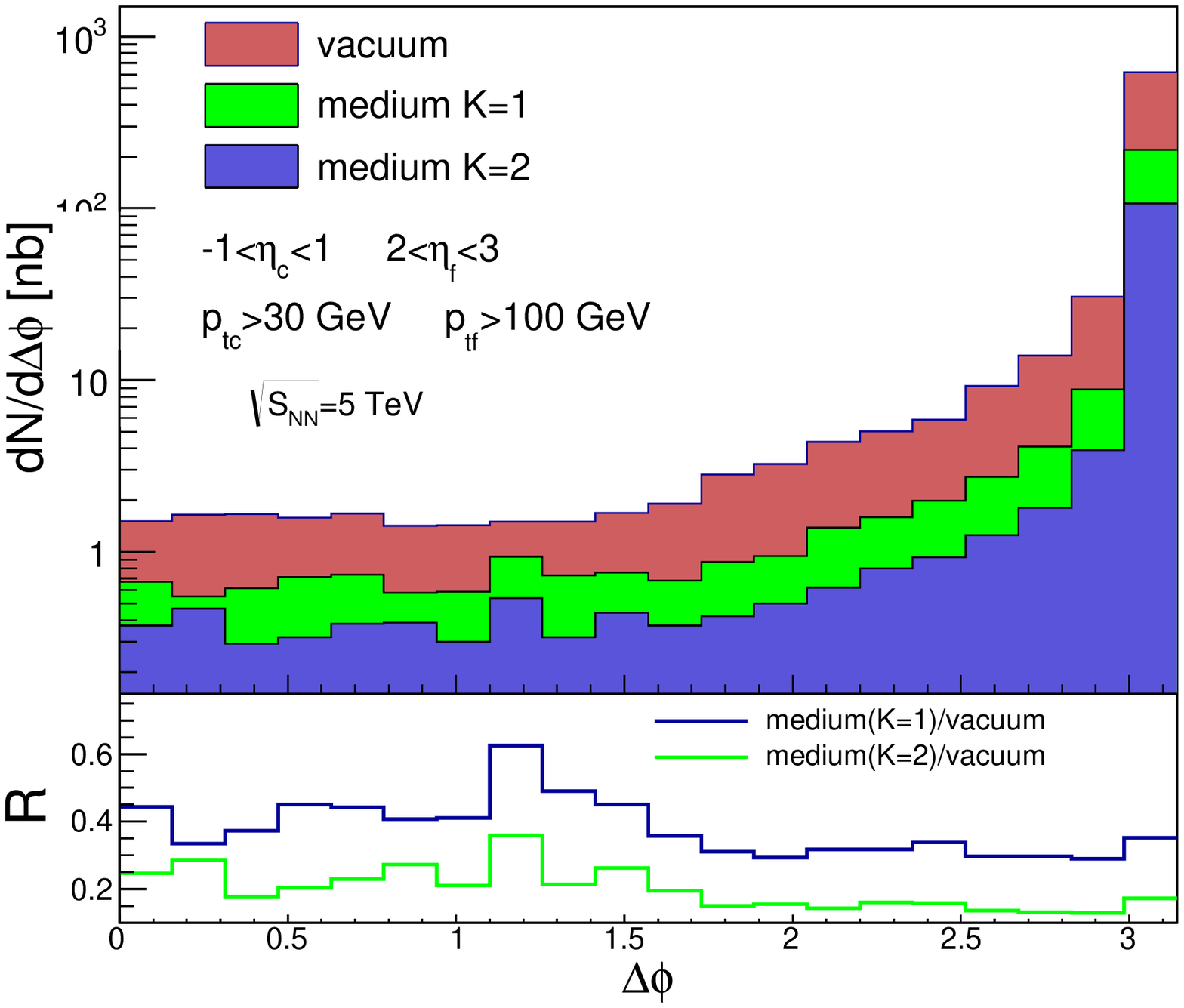}
\caption{}
\label{fig:simplMdphi}
\end{subfigure}
\caption{\small{(a) On top: Rapidity difference between the jets $\Delta\eta$ before and after the jets pass through the medium.  The red histogram represents the $\Delta\eta$ of the jet before passing the medium. The light green histogram represents the $\Delta\eta$ spectrum of a jet quenched by the medium with transport coefficient $K=1$~\eqref{eq:qhat}. The blue histogram represents the $\Delta\eta$ spectrum of a jet quenched by the medium with transport coefficient $K=2$. On bottom: ratios of the histograms. (b) On top: Azimuthal angle between the jets $\Delta\phi$ before and after the jets pass through the medium. The red histogram represents the $\Delta\phi$ of the jet without passing the medium. The dark green histogram represents the $\Delta\phi$ spectrum of a jet quenched by the medium with $K=1$. The green histogram represents the $\Delta\phi$ spectrum of a jet quenched by the medium with $K=2$. On bottom: ratios of the histograms.}}
\end{figure}

\begin{figure}[t!]
\centering
\includegraphics[scale=.5]{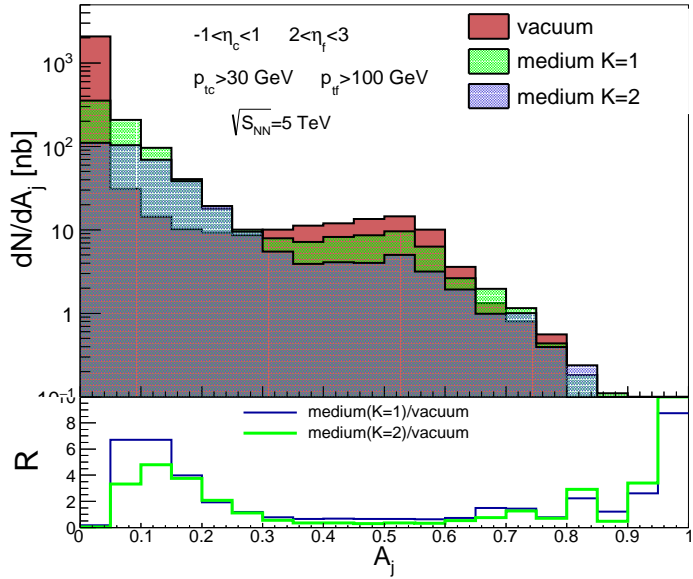}
\caption{\small{On top: Relative transversal momentum difference between the jets $A_j$ before and after the jets passe through the medium. The red histogram represents the $A_j$ of the jet without passing the medium. The green transparent histogram represents the $A_j$ spectrum of a jet quenched by the medium with $K=1$~\eqref{eq:qhat}. The blue transparent histogram represents the $A_j$ spectrum of a jet quenched by the medium with $K=2$. On bottom: ratios of the histograms.}}
\label{fig:simplMAj}
\end{figure}

To outline the workings of the implementation of the model we have chosen five observables: transversal momentum of the central jet ${p_{t}}_c$ (Fig.~\ref{fig:simplMptc}), transversal momentum of the forward jet ${p_{t}}_f$ (Fig.~\ref{fig:simplMptf}), rapidity distance between the jets $\Delta\eta$ (Fig.~\ref{fig:simplMdeta}), azimuthal angle between the jets $\Delta\phi$ (Fig.~\ref{fig:simplMdphi}) and a relative transversal momentum difference of the jets $A_j=\left({p_{t}}_c-{p_{t}}_f\right)/\left({p_{t}}_c+{p_{t}}_f\right)$ (Fig.~\ref{fig:simplMAj}). In each figure we plot the unsuppressed cross section of jets passing through vacuum and two suppressed cross sections of jets passing through nuclear medium with two different nuclear medium parameters ${\hat q}$ modified by choosing the value of the constant $K=1$ and $K=2$. Each of the plots is accompanied by a plot of medium suppression calculated as a ratio of suppressed over  unsuppressed cross sections.

We have chosen the transversal momentum of the central jet is ${p_{t}}_{c}>100\;$GeV. The rapidity of the central jet was $-1<\eta_c<1$. The transversal momentum of the forward jet ${p_{t}}_{f}>30\;$GeV. The rapidity of the forward jet is moderate, $2<\eta_f<3$, in order to be within the reach of current experimental capabilities at LHC.
To evaluate the cross sections, we have used the nCTEQ15FullNuc$\_$208$\_$82~\cite{Kovarik:2015cma} nuclear (lead) PDF for the collinear parton. For the
off-shell gluon density, we have used the novel nuclear TMD PDF constructed applying the KMRW procedure
to the nCTEQ15FullNuc$\_$208$\_$82 collinear set.

We can see in Fig.~\ref{fig:simplMptc} that the suppression by the medium is stronger for lower transversal momenta and gets weaker by increasing ${p_{t}}_c$, which is consistent with other available results in the literature. The same behaviour with suppression decreasing with increasing momentum is present in Fig.~\ref{fig:simplMptf}. The peak in the ${p_{t}}_f$ spectrum at $100\;$GeV corresponds to the back-to-back di-jet configuration. The medium suppression ratio grows with the momentum for ${p_{t}}_f>100\;$GeV, but for ${p_{t}}_f<100\;$GeV the behaviour is the opposite with the ratio growing with decreasing ${p_{t}}_f$. The latter indicates that smaller ${p_{t}}_f$ values are associated with bigger ${p_{t}}_c$ values.

In the rapidity difference spectrum in Fig.~\ref{fig:simplMdeta} we can see that the suppression grows slightly for increasing $\Delta\eta$ as a consequence of the rapidity dependence model of the ${\hat q}$ parameter~\eqref{eq:modelrap}.

The behaviour seen in the $\Delta\phi$ distribution in Fig.~\ref{fig:simplMdphi} is due to the effect of Sudakov resummation. Similar structures have already been been observed in \cite{vanHameren:2014ala}, albeit for at higher rapidities and transverse momenta. This effect arises because of the reshuffling of events from the strictly back-to-back limit to lower momenta, conserving the total number of events. The resulting structure is partly suppressed by medium effects. In the region where $\Delta\phi<1$, for di-jets not passing through the medium, the distribution is completely flat. On the other hand, for di-jets which have passed through the medium, the dependence on $\Delta\phi$ emerges with a distribution slowly falling with decreasing $\Delta\phi$.

The peak, in the Fig.~\ref{fig:simplMAj}, at $A_j\approx(100-30)/(100+30)\approx 0.54$ corresponds to the back-to-back peak in the Fig.~\ref{fig:simplMptf}. Furthemore, as one can see in the ratio plot, the medium suppression is reshuffling di-jets from configurations with ${p_{t}}_c\approx{p_{t}}_f$ to configurations with unequal momenta. This effect becomes stronger after increassing the medium transport coefficient ${\hat q}$ by increasing the constant $K$~\eqref{eq:qhat} from $K=1$ to $K=2$.

\section{Conclusions}
\label{sec:conclusions}

We have proposed a new framework merging HEF with final-state processes in a deconfined medium. The framework allows to investigate the longitudinal structure of QGP and the pattern of decorrelations in QGP that ultimately is the result an interplay of medium effects and vacuum effects. In order to carry out such a task, we have also introduced a new TMD nuclear PDF for a realistic modeling of initial-state nuclear effects. 

The proposed framework could serve to disentangle effects related to energy loss from those related to angular decorrelation whether due to initial-state/saturation or final-state broadening in the quark-gluon plasma. It also allows to calculate observables that are potentially sensitive to physics at forward rapidity; in particular, the longitudinal structure of the plasma. We have calculated distributions involving cuts and parameters realistic for heavy-ion experiments at the LHC.

Our study confirms that the bulk component of the decorrelations is due to a vacuum initial-state shower. This is mainly because medium-induced energy loss mainly shifts the $p_t$-spectra of the outgoing jets. The presence of medium interactions changes the normalization and to some extend shape of distributions of the studied observables. This is a clear prediction from this particular model, and would be interesting to compare with experimental data. 

We are currently limiting ourselves to high-energy processes, where medium-modifications factorise from the hard cross section and affect mainly the resulting $p_t$-distributions of the outgoing jets.
In the future we plan to study more forward processes and therefore to generalize the framework to account for saturation effects. We also plan to study the impact of jet substructure fluctuations, in the spirit of \cite{Mehtar-Tani:2017ypq} that are crucial for describing high-$p_t$ data at mid-rapidity \cite{Milhano:2015mng}. Furthermore, in order to shed more light on the role of final-state broadening, we plan to implement angular deflection due to final-state momentum broadening in the spirit of \cite{Mueller:2016gko,Mueller:2016xoc}, which can play a role at lower colliding energies. This could give rise to a more intricate pattern of medium-induced modifications.


\section*{Acknowledgments}
The work of M.D. and K.K. was supported by Narodowe Centrum Nauki
with Sonata Bis grant DEC-2013/10/E/ST2/00656.
K.T. has been supported by a Marie Sklodowska-Curie Individual Fellowship of the European Commission's Horizon 2020 Programme under contract number 655279 ResolvedJetsHIC. 
We acknowledge discussions with Andreas van Hameren on the details of implementation of medium effects into the KaTie event generator and discussions with Doga Can Gulhan on the experimental aspects of production of forward jets in Pb-Pb collision. Figure 1 was made with the JaxoDraw package \cite{Binosi:2008ig}.

\end{document}